\documentclass{jpsj3}
\usepackage{graphicx,amsmath,amssymb,bm}

\allowdisplaybreaks 

\usepackage{color}
\definecolor{Green}{rgb}{0,0.7,0}

\newcommand{\bk}{\bm{k}}

\newcommand{\tx}{\bk \cdot \bm{a}}
\newcommand{\ty}{\bk \cdot \bm{b}}
\newcommand{\tz}{\bk \cdot \bm{c}}
\newcommand{\p}{ $+1$}
\newcommand{\m}{$-1$}

\begin{document}


\title{
A tight-binding model of an ambient-pressure molecular Dirac electron system 
 with open nodal lines 
}
\author{
Reizo \textsc{Kato}$^{1}$
\thanks{E-mail: reizo@riken.jp}
 and 
Yoshikazu \textsc{Suzumura}$^{2}$
}
\inst{
$^1$
RIKEN, 2-1 Hirosawa, Wako-shi, Saitama 351-0198, Japan \\
$^2$
Department of Physics, Nagoya University,  Chikusa-ku, Nagoya 464-8602, Japan \\
}

\recdate{January 13,2020}
\abst{
By deriving a tight-binding model, we demonstrate a mechanism of forming 
  a nodal line of Dirac points in a single-component molecular conductor
  [Pt(dmdt)$_2$] [Zhou {\it et al.}, Chem. Commun. {\bfseries 55}, 3327 (2019)], 
  consisting of HOMO and LUMO. 
The nodal line is obtained as the intersection of two surfaces, where 
one corresponds to the HOMO-LUMO band crossing and another 
 is vanishing of the HOMO-LUMO couplings due to their different symmetries.  
 The latter property is essential for the Dirac electron in molecular conductors.  
The nature of the open nodal line is discussed in terms of the parity of the wavefunctions at eight TRIMs (time reversal invariant momenta).
}


\maketitle
\section{Introduction}
The Dirac electron system has attracted a great deal of attention from theoretical and experimental viewpoints.\cite{Novoselov_Nature}
  A variety of materials with Dirac points in energy band structure have been developed.
\cite{Hirayama2018,Bernevig2018} 
However, the number of systems where the Dirac point is located in the vicinity of Fermi level is still limited. 
The Dirac electron systems in molecular conductors\cite{Kajita2014} 
have been extensively studied, since the zero-gapped state with the Fermi level at the Dirac point was found in a two-dimensional molecular conductor, $\alpha$-(BEDT-TTF)$_2$I$_3$ (BEDT-TTF=Bis(ethylenedithio)tetrathiafulvalene).\cite{Katayama2006_JPSJ75} 
Among them, we found a nodal line semimetal state in a single-component molecular conductor based on a metal dithiolene complex [Pd(dddt)$_2$] (dddt=5,6-dihydro-1,4-dithiin-2,3-dithiolate) under high pressure.\cite{Kato_JACS}
 At ambient pressure, [Pd(dddt)$_2$] is a normal band insulator with fully occupied HOMO band and completely empty LUMO band. Since the metal dithiolene complexes with the square planar coordination geometry have a small HOMO-LUMO energy gap, enlargement of the bandwidth by the application of pressure can induce overlapping energy bands. Indeed, under pressure, the HOMO band and the LUMO band with opposite curvatures in [Pd(dddt)$_2$] overlap and induce
 electron transfer from the HOMO band to the LUMO band. 
The node of the HOMO-LUMO coupling, where the gap formation does not work, provides the Dirac point that is located around the Fermi level. The mechanism of the Dirac cone formation in this system can be understood using an effective model of 2 $\times$ 2 Hamiltonian.\cite{Kato2017_JPSJ,Tsumuraya2018_JPSJ,Liu2018} 
Within the three-dimensional Brillouin zone, the Dirac point moves in a loop. 
The topological number indicates that the system is a topological nodal line semimetal.

These results indicate that single-component molecular conductors can easily provide such a type of Dirac electron system. 
Indeed, after our work, an ambient-pressure Dirac electron system based on a single-component molecular conductor [Pt(dmdt)$_2$] 
(dmdt=dimethyltetrathiafulvalenedithiolate) was disclosed by the first principles DFT (Density Functional Theory) band calculation.\cite{Zhou2019}
 We report here a tight-binding model for [Pt(dmdt)$_2$]  and  indicate that this system demonstrates a typical and simple example of the Dirac cone formation mechanism in the single-component molecular conductors.
 We also discuss the nature of the open nodal line in terms of the parity of the wavefunctions at eight TRIMs (time reversal invariant momenta).

\section{Model}

\subsection{Crystal structure and intermolecular transfer integrals} 
\begin{figure}
  \centering
\includegraphics[width=7.5cm]{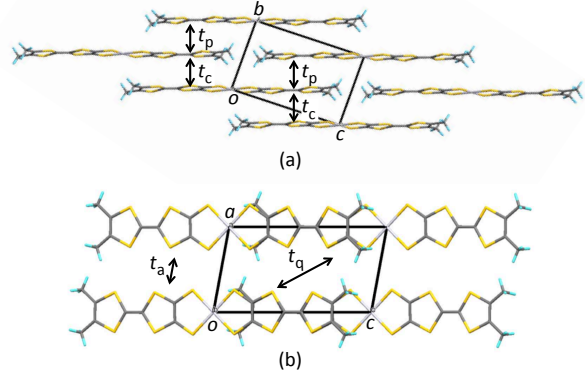}   
  \caption{(Color online)
Crystal structure of [Pt(dmdt)$_2$] 
and intermolecular couplings.
}
\label{fig1}
\end{figure}

Crystal structure of [Pt(dmdt)$_2$] is very simple.\cite{Zhou2019}
 The space group is $P\bar{1}$ and the unit cell 
 ($a=6.620$\AA, $b$=7.611\AA, $c$=11.639\AA, 
$\alpha$=86.05$^\circ$, 
$\beta$=78.98$^\circ$, 
$\gamma$=75.04$^\circ$)
 contains only one molecule on the inversion center. 
Molecular packing of [Pt(dmdt)$_2$] units within the $b$--$c$ plane exhibits 
the stretcher bond arrangement with the face-to-face overlap between molecules of half a molecule, which forms a two-dimensional (2D) conduction layer 
(Fig.~\ref{fig1}(a)). 
Along the interlayer direction (parallel to the $a$ axis), [Pt(dmdt)$_2$] 
molecules are arranged uniformly in the side-by-side manner
(Fig.~\ref{fig1}(b)).  

Since the unit cell contains one [Pt(dmdt)$_2$] molecule with HOMO and LUMO, we consider one HOMO band and one LUMO band in the tight-binding model. 
 Calculations of molecular orbitals and intermolecular overlap integrals 
($S$) between HOMOs and LUMOs by extended H\"uckel method 
 were carried out on the basis of crystal structure 
 data.~\cite{Zhou2019} 
Reported sets of semi-empirical parameters for Slater-type atomic orbitals and valence shell ionization potentials 
for H,~\cite{Hoffmann_1} C,~\cite{Hoffmann_1} S,~\cite{Clementi} and 
Pt~\cite{Hoffmann_2} were used for the calculations.
 Intermolecular transfer integrals, $t$ (eV), 
were estimated using the equation $t = - 10S$ (Table~\ref{table1}). 
Intralayer transfer integrals ($t_c$, $t_p$) 
including HOMO-LUMO couplings 
are found to be about 10 times greater than interlayer ones ($t_a$, $t_q$), 
 which indicates that the system has a 2D network parallel 
 to the $b$--$c$ plane with small but significant interlayer interactions.
 An important point is that the transfer integrals for intralayer HOMO-HOMO and LUMO-LUMO couplings ($t_c$, $t_p$) have opposite signs 
 and thus the HOMO band is convex upward and the LUMO band is convex downward.

\begin{table}
\caption{ 
 HOMO-HOMO (H), LUMO-LUMO (L), and HOMO-LUMO (HL) intermolecular transfer Integrals (meV)\\
 }
\begin{center}
\begin{tabular} {cccc}
\hline\noalign{\smallskip}
 Transfer integral*       & H      &  L    & HL     \\
\hline\noalign{\smallskip}
 Intralayer \;\;$t_c$      & $67.1$   &  $-62.9$   & $64.9$    \\
 \;\; \;\;\;\;\;\;\;\;\;\;\;\; \;\;$t_p$      & $53.4$    &  $-49.8$   & $51.7$     \\
\hline\noalign{\smallskip}
 Interlayer \;\; $t_a $       & $-6.2$   &  $-6.5$   & $0.3  $      \\
 \;\; \;\;\;\;\;\;\;\;\;\;\;\; \;\;$t_q $       & $8.2$    &  $-7.4$   & $-7.8  $        \\
\noalign{\smallskip}\hline
\end{tabular}
\end{center}
* See Fig.~1. Transfer integrals in this table are used in Eqs.~(2a)-(2c) with subscripts H, L, and HL, that represent HOMO-HOMO, LUMO-LUMO, and HOMO-LUMO couplings, respectively (for example, 
$t_{\rm cH}$ means a transfer integral $t_{\rm c}$ between HOMO and HOMO).

\label{table1}
\end{table}


\subsection{Tight-binding model and energy band}
Using transfer integrals in Table~\ref{table1}, the band energies $E$($\bk$) 
($\bk$ is given by  
$\bk = k_x\bm{a}^* + k_y \bm{b}^* + k_z \bm{c}^*$ in terms of 
the reciprocal lattice vectors $\bm{a}^*$, $\bm{b}^*$, and $\bm{c}^*$)
 are obtained as eigenvalues of the following simple 2 $\times$ 2 Hermitian matrix.
\begin{eqnarray}
 {\bf H}(\bk)  
 &=& 
\begin{pmatrix}
 h_{\rm H}  &  h_{\rm HL}  \\
 \overline{h_{\rm HL}}  &  h_{\rm L}
\end{pmatrix} \ , 
\label{eq:eq1}
\end{eqnarray}
where the base is taken as HOMO and LUMO, and 
the matrix elements are given by 

\begin{subequations}
\begin{eqnarray}
\label{eq:eq2a}
h_{\rm H} &=&  2 [t_{\rm cH} \cos(\tz) +  t_{\rm aH} \cos(\tx) 
      \nonumber \\
   &+ &  t_{\rm pH} \cos(\ty+\tz) +  t_{\rm qH} \cos(- \tx -\tz)] 
  \; ,  \\
\label{eq:eq2b}
h_{\rm HL} &=& 2i [t_{\rm cHL} \sin(\tz) +  t_{\rm aHL} \sin(\tx) 
      \nonumber \\
     &+&   t_{\rm pHL} \sin(\ty + \tz) +  t_{\rm qHL} \sin(-\tx-\tz)] 
                 \nonumber  \\
& \equiv & i f(\bk) \; ,
   \\
\label{eq:eq2c}
h_{\rm L}&=&    2[ t_{\rm cL} \cos(\tz) +  t_{\rm aL} \cos(\tx) 
      \nonumber \\
   & +&  t_{\rm pL} \cos(\ty+\tz) +  t_{\rm qL} \cos(- \tx -\tz)] 
 +  \Delta  
\; . 
\end{eqnarray}
\end{subequations}
The energy gap between HOMO and LUMO was taken as $\Delta$ = 0.25 eV 
to reproduce the results of the first-principles calculation 
in Ref.~\citen{Zhou2019}. 
Matrix elements $h_{\rm H}$, $h_{\rm L}$, and $h_{\rm HL}$ 
are associated with HOMO-HOMO, LUMO-LUMO, and HOMO-LUMO couplings, respectively.  The quantity $f(\bk)$
in  Eq.~(\ref{eq:eq2b}) is defined  for the later discussion.
   
\begin{figure}
  \centering
\vspace{0.5cm}
\includegraphics[width=13cm]{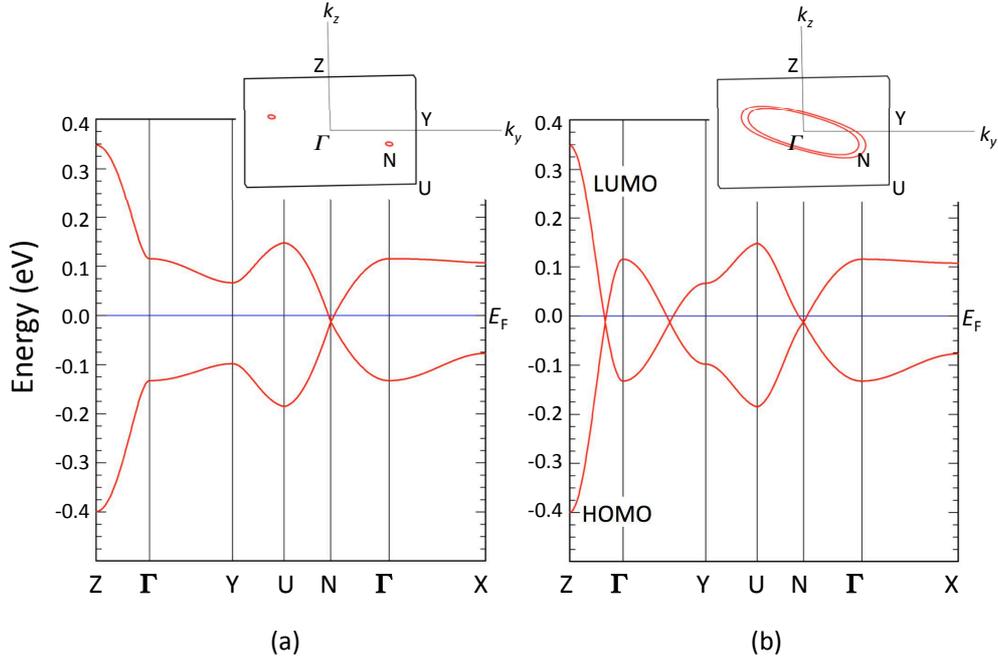}   
\caption{(Color online)
Band dispersion and Fermi surface ($k_x =0$)
(a) with HOMO-LUMO couplings  (pristine)
(b) without HOMO-LUMO couplings
 $(|h_{\rm HL}| = 0)$.
N $(0, 0.334, -0.118)$ denotes a location of the Dirac point.
}
\label{fig2}
\end{figure}

\begin{figure}
  \centering
\includegraphics[width=12cm]{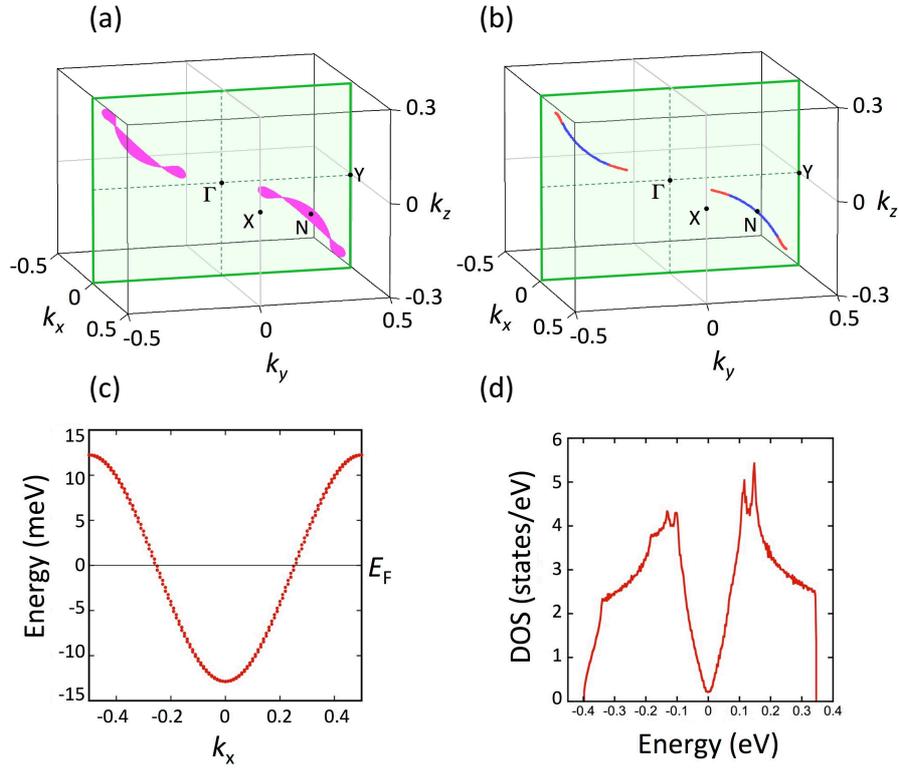} 
  \caption{(Color online)
Three-dimensional plot of (a) Fermi surface and (b) nodal line, where the hole-like character is indicated in red and the electron-like character in blue. (c) $k_x$  dependence of the band energy at the Dirac point. (d) Density of states (DOS) per unit cell. All of them correspond to the pristine energy band structure shown in Fig.~\ref{fig2}(a). 
}
\label{fig3}
\end{figure}

\begin{figure}
  \centering
\includegraphics[width=12cm]{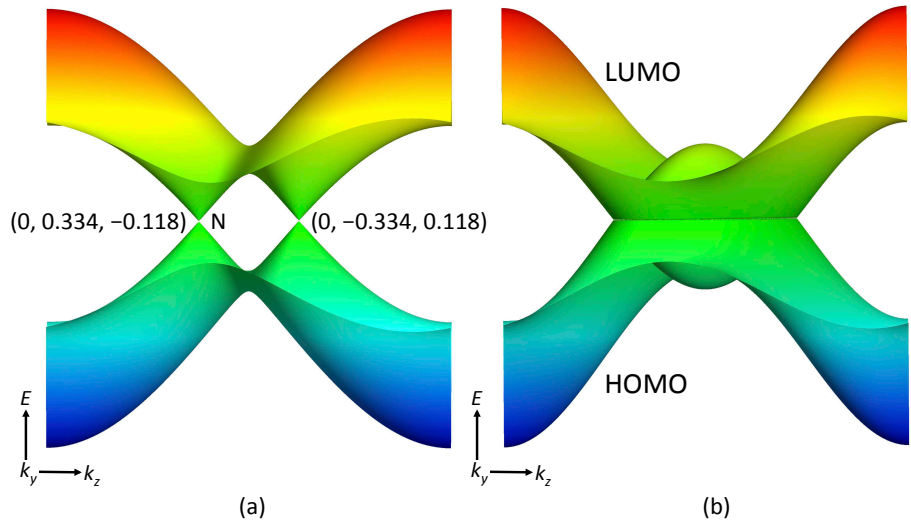}   
  \caption{(Color online)
Band energy dispersion surface at $k_x=0$ 
(a) with HOMO-LUMO couplings (pristine)
(b) without HOMO-LUMO couplings $(|h_{\rm HL}| = 0)$.
These show the difference in the energy band between 
Figs.~\ref{fig2}(a) and (b).
 }
\label{fig4}
\end{figure}

Figures \ref{fig2}(a) and \ref{fig3}(a) show band dispersion 
and Fermi surface of the tight-binding model. 
A band crossing occurs at 
$(k_x,k_y,k_z) = (0, \pm 0.334, \mp 0.118)$ 
on the $k_x=0$ plane and 
a pair of Dirac points located symmetrically 
to the $\Gamma$ point emerge (Fig.~\ref{fig4}(a)). 
On the other hand, the Dirac points 
appear at $(0, \pm 0.381, \mp 0.202)$ 
in the DFT calculation. 
The energy at the Dirac points is slightly lower than the Fermi level, 
  which indicates the electron-like character of the Fermi surface. 
Due to the interlayer interactions, the Dirac point draws a wavy curve
 along the $k_x$ direction in the reciprocal lattice (Fig.~\ref{fig3}(b)). 
Consequently, in contrast to the case of [Pd(dddt)$_2$] with the looped nodal line, 
the present system has a pair of open nodal lines.
On the $k_x =0.5$ plane, 
the Dirac points emerge at $(0.5, \pm 0.305, \mp 0.141)$, 
while at $(0.5, \pm 0.326, \mp 0.160)$ 
in the DFT calculation. Depending on $k_x$, 
the band energy $E$ (eV) at the Dirac point changes as shown in Fig.~\ref{fig3}(c) 
and the $k_x$ dependence can be described as 
$E(k_x) = - 0.125 \cos(2 \pi k_x)$. 
Figure \ref{fig3}(c) indicates that the electron-like Fermi surface turns to the hole-like one at $k_x=\pm 0.25$ and 
 explains surviving density of states (DOS) at the Fermi level 
 in Fig.~\ref{fig3}(d). 
The linear energy dispersion in the DOS around the Fermi level in the range of ca. $\pm$0.1eV is closely similar to the result of the DFT calculation. 
We note the energy dispersion  of the Dirac cone,
 which is given  on the $k_y$--$k_z$ plane with the fixed $k_x$.
 The cross section close to the Dirac point shows an ellipse,
 with a velocity $V_-$ ($V_+$) for the  major (minor) axis, which is
 rotated by
 an angle $\theta$ from the $k_y$ ($k_z$) axis.
  For $k_x$ = 0 (0.5), we obtain
   $V_\pm \simeq  C_\pm \times 10^{5}$ (m/s),
 with $C_+$ = 2.1 (2.2), $C_-$ = 1.5 (1.2), and
$\theta$ = $-25^\circ$ ($-36^\circ$).
Thus, the anisotropy and the $k_x$ dependence of the cone
 are moderately visible, while  the velocity of the tight-binding
 model  is slightly
 smaller than that of the DFT calculation
 with  the maximum velocity
 $\simeq 4 \times 10^{5}$ (m/s).~\cite{Zhou2019}

All these results well reproduce the essential feature of the energy band structure obtained by the DFT calculations and indicate that this system is a nodal line semimetal.

\subsection{Dirac cone formation and nodal line}
\begin{figure}
  \centering
\includegraphics[width=12cm]{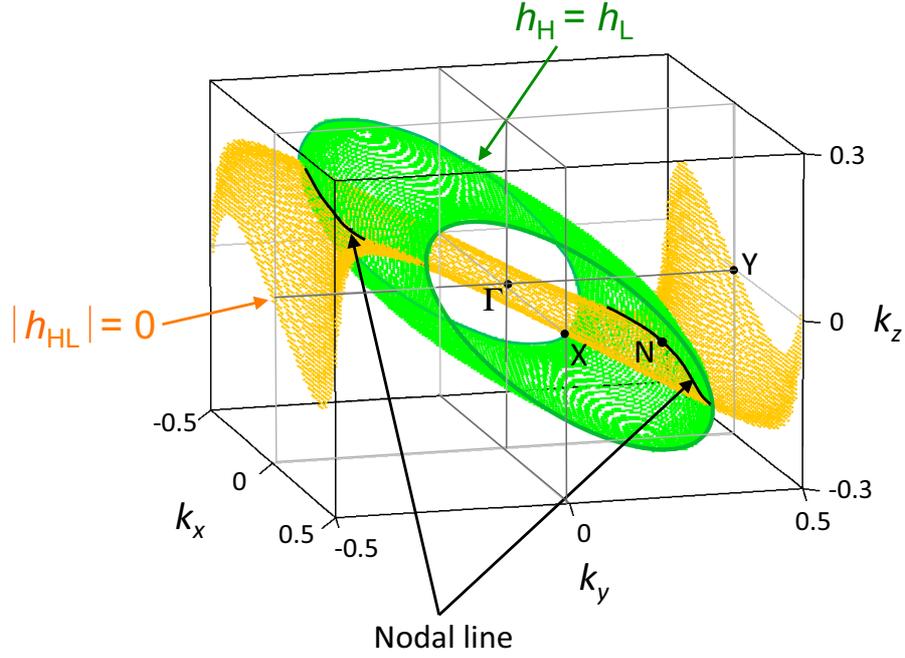}   
  \caption{(Color online)
Intersection of HOMO and LUMO bands ($h_H = h_L$),
 nodal surface on which $|h_{\rm HL}|$ = 0, 
and nodal line (see text). 
 }
\label{fig5}
\end{figure}

We now discuss the Dirac cone formation in this single-component molecular conductor. The Dirac point emerges under the following conditions:
 \begin{eqnarray}
  && h_{\rm H} =  h_{\rm L}  \; , 
 \label{eq:eq3} \\
  & & |h_{\rm HL}| = 0  \; .
 \label{eq:eq4}
\end{eqnarray}
Equation (\ref{eq:eq3}) corresponds 
 to the HOMO-LUMO band crossing, and 
Eq.~(\ref{eq:eq4}) gives a node of the HOMO-LUMO coupling. 
When there is no HOMO-LUMO coupling, the HOMO and LUMO bands cross
 each other (Figs.~\ref{fig2}(b) and \ref{fig4}(b)) and Eq.~(\ref{eq:eq3}) 
 gives the intersection of the HOMO and LUMO bands as a distorted cylinder (green surface in Fig.~\ref{fig5}). 
Since the HOMO band is fully occupied and the LUMO band is completely empty 
 originally, the intersection is located in the vicinity of the Fermi level. 
This is an important and general feature of the single-component molecular conductor. 
In general, an introduction of the HOMO-LUMO coupling removes the degeneracy at the intersection and opens a gap. 
This means that the metallic state turns to a semiconducting one 
 due to the HOMO-LUMO coupling. 
In this system, however, there exist surfaces 
 on which the HOMO-LUMO coupling that opens the gap is zero 
(orange surface in Fig.~\ref{fig5})
  given by Eq.~(\ref{eq:eq4}). 
The intersection of these two surfaces that satisfies 
 $h_{\rm H} = h_{\rm L}$ 
and $|h_{\rm HL}| = 0$ provides the nodal line (black line in Fig.~\ref{fig5})
 on which the conduction and valence bands touch each other 
 at one point with linear dispersion. 
Here we note a property of  $h_{\rm HL}$, i.e., 
  $f(\bk)$ in Eq.~(\ref{eq:eq2b}). 
Since $f(\bk)$ is real and satisfies $f(\bk) = -f(-\bk)$
 owing to a different symmetry of  
  HOMO and LUMO,\cite{Tsumuraya2018_JPSJ}
 we obtain $f(0) = 0$.
For small $\tx (= 2\pi k_x), \ty (=2\pi k_y)$ 
 and $\tz (=2\pi k_z)$, $f(\bk)$ can be expressed  as 
 a linear combination of $k_x$, $k_y$ and $k_z$, i.e.,
$f(\bk)/(4\pi)  \simeq$  $ (t_{\rm aHL}-t_{\rm qHL})k_x 
 +  t_{\rm pHL} k_y
 + (t_{\rm cHL}-t_{\rm qHL}+t_{\rm pHL})k_z$. 
Thus,  the plane determined by $f(\bk)=0$ includes the $\Gamma$ point and 
 becomes  almost perpendicular to the $k_y$--$k_z$ plane 
 for small interlayer transfer energies.

\begin{figure}
  \centering
\includegraphics[width=5cm]{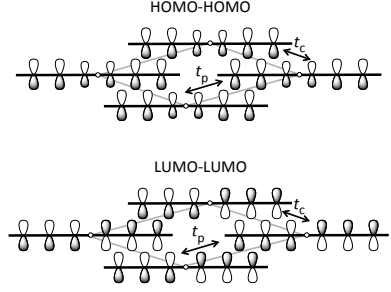}   
  \caption{
HOMO-HOMO and LUMO-LUMO couplings in the stretcher bond arrangement
 of the [Pt(dmdt)$_2$] molecule. Each frontier molecular orbital is represented using the p orbital of sulfur atom that governs intermolecular transfer integrals. 
 }
\label{fig6}
\end{figure}

Requirements for the emergence of the single-component molecular
  Dirac electron system are 1) HOMO-LUMO band crossing and 
2) node of the HOMO-LUMO couplings.
 In order to satisfy requirement 1), main HOMO-HOMO and LUMO-LUMO transfer 
integrals should have opposite signs. 
This depends on the molecular arrangement associated with the symmetry 
 of the frontier molecular orbitals. 
From this viewpoint, the stretcher bond arrangement in the conduction layer 
 of the present crystal (Fig.~\ref{fig1}(a)) is quite suitable for the formation of 
 the crossing band structure. 
In the metal dithiolene complex molecule [Pt(dmdt)$_2$],  
  HOMO has ungerade (odd) symmetry 
 and LUMO has gerade (even) symmetry. 
Therefore, as shown in Fig.~\ref{fig6}, 
the HOMO-HOMO and LUMO-LUMO couplings give overlap 
 and transfer integrals ($t_c$ and $t_p$)  with opposite signs 
in the stretcher bond arrangement, where each ligand overlaps 
 with the ligand on the opposite side of the adjacent molecule. 

\subsection{Open nodal line and parity}

\begin{figure}
  \centering
\includegraphics[width=10cm]{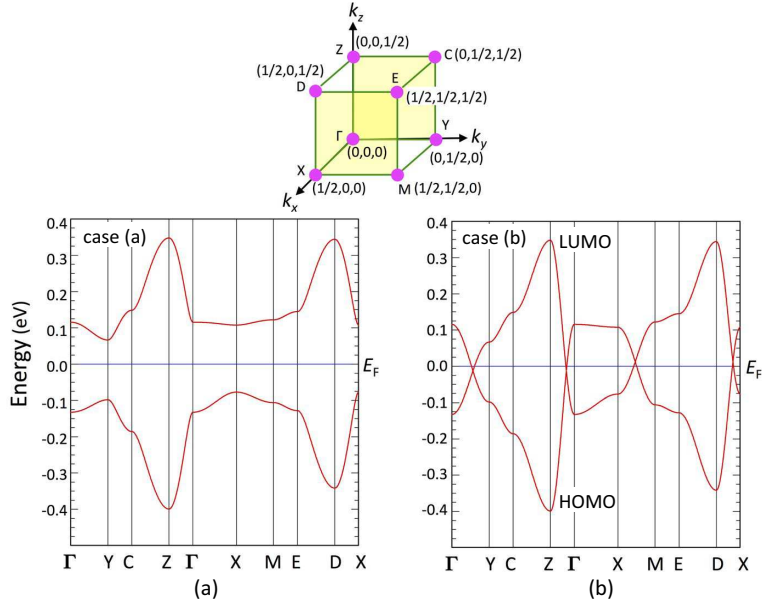}   
  \caption{
Eight TRIMs in the Brillouin zone, which are associated with  
the parity in Table \ref{table_2}. The energy bands connecting TRIMs 
for the cases (a) and (b) are compared. 
 }
\label{fig7}
\end{figure}

 We  analyze the open nodal line  
using  the parity    at the TRIMs ($\bm{k}= \bm{G}/2$ 
      with $\bm{G}$ being the reciprocal lattice vector). 
From Eq.~(\ref{eq:eq1}),  energy $E_j$ and wavefunction $\Psi_j$ ($j$=1, 2) 
at the TRIMs are calculated as    
\begin{eqnarray}
 \bm{H}(\bm{G}/2) \Psi_j(\bm{G}/2) = E_j(\bm{G}/2)\Psi_j(\bm{G}/2) \; ,
 \label{eq:TRIM}
\end{eqnarray}
 where  $E_1(\bm{G}/2) > E_2(\bm{G}/2)$.
   The parity is obtained  from  a spatial inversion 
 for  $\Psi_j(\bm{G}/2)$,~\cite{Piechon2013} 
     using  the inversion center
       on the Pt atom at a lattice point.   
    Such a 2 $\times$ 2 inversion matrix 
      $\hat{P}_{\rm I}$  has only  diagonal elements given by 
        $- 1$ and $+1$ for the base of HOMO and LUMO,  respectively.
Since  $[\bm{H}(\bm{G}/2), \hat{P}_{\rm I}(\bm{G}/2)]=0$,
  both $\bm{H}(\bm{G}/2)$ and  $\hat{P}_{\rm I}(\bm{G}/2)$ have a common
 eigenfunction. 
 Thus,   we obtain
     $ \hat{P}_{\rm I}(\bm{G}/2) \Psi_j(\bm{G}/2)$
       = $E_P(j,\bm{G}/2) \Psi_j(\bm{G}/2)$, ($j$=1 and 2), where  
     $E_P(j,\bm{G}/2) = \pm 1$.
 In terms of $\Psi_j(\bm{G}/2)$, the parity eigenvalue $E_P(j,\bm{G}/2)$
  is  calculated as, 
\begin{eqnarray}
 E_P(j,\bm{G}/2) &=& 
    \Psi_j(\bm{G}/2)^\dagger \hat{P}_{\rm I}(\bm{G}/2 ) \Psi_j(\bm{G}/2) 
  \; , 
 \label{eq:parity_c}
\end{eqnarray}
where 
 $\bm{G}/2$ = (0, 0, 0) ($\Gamma$), (1/2, 0, 0) (X), (0, 1/2, 0) (Y),
  (1/2, 1/2, 0) (M), (0, 0, 1/2) (Z), (1/2, 0, 1/2) (D), 
 (0, 1/2, 1/2) (C), and  (1/2, 1/2, 1/2) (E).
 The parity $ E_P(j,\bm{G}/2)$ corresponding to 
$E_j(\bm{G}/2)$  is summarized in Table \ref{table_2} for 
 both Eq.~(\ref{eq:eq1}) with HOMO-LUMO couplings (pristine, case (a)) and 
 that without  HOMO-LUMO couplings ($|h_{\rm HL}| = 0$, case (b)).
 Here we note a relation between the parity $E_P [= E_P(j,\bm{G}/2)]$ 
and these two types of energy bands 
shown in Figs.~\ref{fig2}, ~\ref{fig4}, and ~\ref{fig7}. 
Figures \ref{fig7}(a) and (b) show the energy bands connecting eight TRIMs  
for the cases (a) and (b), where 
 the band crossing between HOMO and LUMO bands in the case (b) 
 is removed by the HOMO-LUMO couplings in the case (a) except for the nodal line (not shown here). 
Noting that the HOMO (LUMO) band provides  $E_P= -1 ( +1)$, 
 the case (b) in Table \ref{table_2} suggests that 
 $E_1(\bk)$  shows  the  HOMO ($E_P= -1$) band for the TRIMs of $\Gamma$ 
 and X, 
 and the LUMO ($E_P= +1$) band for the rest of  TRIMs.
 Since such a property remains unchanged even for the case (a), 
  the band with the same parity is connected in the presence of  
  the HOMO-LUMO couplings. Then, such a possible distinction  of the $E_P= +1$ and $E_P= -1$ bands is useful for understanding the open  nodal line clearly.

\begin{table}
\caption{ 
Parity eigenvalue $E_{P}(j,\bm{G}/2) (= \pm 1)$ 
 with  the HOMO-LUMO coupling (pristine, case (a))
 and without the HOMO-LUMO coupling ($|h_{\rm HL}| = 0$, case (b))
 which corresponds to Figs.~\ref{fig2} and ~\ref{fig4}, respectively. 
  $E_{P}(j,\bm{G}/2) (= \pm 1)$ corresponds to 
 the eigenvalue of  $E_j(\bm{G}/2)$ at the TRIMs($\bm{G}/2$) 
  shown in Fig.~\ref{fig7}.
}
\begin{center}
\begin{tabular} {cccccccccc}
\hline\noalign{\smallskip}
$E_P(j,\bm{G}/2)$ & & $\Gamma$ & Y& C & Z & X & M & E &D \\
\noalign{\smallskip}\hline\noalign{\smallskip}
case (a) & $j$=1  & \m  & \p  & \p & \p   & \m & \p & \p & \p \\
         & $j$=2  & \p  & \m  & \m & \m   & \p & \m & \m & \m \\
\noalign{\smallskip}\hline\noalign{\smallskip}
case (b) & $j$=1  & \m  & \p  & \p & \p   & \m & \p & \p & \p \\
         & $j$=2  & \p  & \m  & \m & \m   & \p & \m & \m & \m \\
\noalign{\smallskip}\hline
\end{tabular}
\end{center}
\label{table_2}
\end{table}
 
Since there is a relation $\sum_{j} E_P(j,\bm{G}/2)$ = 0 
 for the respective $\bm{G}$,  
  the condition 
for the Dirac point is written either by the occupied band or by  
the empty band.
In terms of  the empty band,
 the explicit form of the condition~\cite{Fu2007_PRB76,Piechon2013} 
 is given by    
\begin{subequations}
\begin{eqnarray}
P &=&  P_0 P_1= \pm 1  \; ,
 \label{Dirac_P}
\end{eqnarray}
where 
\begin{eqnarray}
 P_0  & =&  E_P(1,\Gamma) E_P(1,Y) E_P(1,C) E_P(1,Z) \; ,
\label{Dirac_P_0} 
   \\  
 P_1  & =&  E_P(1,X) E_P(1,M) E_P(1,E) E_P(1,D) \; .
  \label{Dirac_P_1}  
\end{eqnarray}
\end{subequations}

TRIMs of $P_0$ are located on  the plane of $k_x=0$, while 
those of $P_1$ are located  on the plane of $k_x=0.5$ (Fig. 7). 
The quantity $P$, which denotes a  product of 
 parity eigenvalues  of all the empty bands, 
 provides an odd (or even)  number of pair of Dirac points.
The present case is obtained as  follows.
When $P = +1$,  we obtain 
 open nodal line 
($P_0 = -1$ and $P_1 = -1$)
  or the absence of Dirac point 
 ($P_0 = +1$  and $P_1 = +1$ ). 
The former case  corresponds to the present case of [Pt(dmdt)$_2$].
In fact, the nodal lines existing along the $k_x$ axis 
pass through the plane of $k_x$=0
 and $\pm  0.5$. 
Note that  $P = -1$  ($P_0 = \pm 1$, $P_1= \mp 1$) 
 gives the closed nodal line (i.e., loop).
The loop (closed line), 
  which is found for [Pd(dddt)$_2$]\cite{Kato2017_JPSJ},  
  corresponds to  $P = -1$ 
 with $P_0 = -1$  and $P_1= +1$, where $P_0$ ($P_1$)  
 consists of $\Gamma$, X, Y, and M (Z, D, C, and E), respectively.
Thus, the analysis in term of the parity provides 
 a clear distinction between  the open  and closed nodal lines.

\section{Summary}
We have studied the ambient-pressure molecular Dirac electron system in the single-component crystal of the metal dithiolene complex [Pt(dmdt)$_2$] using the tight-binding model based on the extended H\"uckel molecular orbital calculations. 
The model well reproduces the essential feature of the results of first-principles DFT calculations including the emergence of the Dirac cones. 
Opposite signs of intermolecular HOMO-HOMO and LUMO-LUMO transfer integrals are attributed to the stretcher bond arrangement of [Pt(dmdt)$_2$] molecules in the conduction layer, and lead to the HOMO-LUMO band crossing. 
The intersection of the HOMO and LUMO bands forms a distorted cylinder in the Brillouin zone, when there is no HOMO-LUMO coupling. 
The HOMO-LUMO coupling vanishes on the plane containing the $\Gamma$ point. 
The Dirac point emerges at the point where these two surfaces meet, and describes the open nodal line. 
The nature of the open nodal line can be analyzed in terms of the parity of the wavefunctions at eight TRIMs. 
All these results obtained by the tight-binding model demonstrate that [Pt(dmdt)$_2$] is a typical and simple example of the Dirac electron system based on the single-component molecular conductors and promise the existence of next materials having similar electronic structures.

\acknowledgements

This work was supported 
 by JSPS KAKENHI Grant,  JP16H06346.


\end{document}